\documentclass[twocolumn]{article} 
\usepackage{latex8}
\usepackage{times}
\usepackage{nopageno}
\usepackage[dvips]{epsfig}
\def\FT{Fault-\kern-2pt Tolerant}
\def\WWW{World-\kern-1pt Wide Web}

\def\nmr{\hbox{$N$\kern-1pt MR}}
\def\taskd{task \hbox{$\mathcal{D}$}}
\def\taski{task \hbox{$\mathcal{I}$}}
\def\taskr{task \hbox{$\mathcal{R}$}}
\def\taska{task \hbox{$\mathcal{A}$}}
\def\Taska{Task \hbox{$\mathcal{A}$}}
\def\mod{\hbox{mod}}

\font\sevenrm=cmr7
\font\nineit=cmti9
\let\mc=\ninerm 
\newbox\PPbox 
\setbox\PPbox=\hbox{\kern.5pt\raise1pt\hbox{\sevenrm+\kern-1pt+}\kern.5pt}
\def\PP{\copy\PPbox}
\def\CPLUSPLUS{{\mc C\kern-1pt\PP\spacefactor1000}}
\def\ICPLUSPLUS{{\nineit C\kern-1pt\mc\PP\spacefactor1000}}
\def\UNDSCR{\hskip-1pt\_}
\begin{document}
\thispagestyle{empty}
%
\title{An Algorithm for Tolerating Crash Failures in Distributed Systems}
%
\author{Vincenzo De Florio, Geert Deconinck, and Rudy Lauwereins}
%
\affiliation{Katholieke Universiteit Leuven, \\
Electrical Engineering Department, ACCA Group, \\
Kard. Mercierlaan 94, B-3001 Heverlee, Belgium.}
\email{$\{$vincenzo.deflorio$|$lauwerin$\}$@gmail.com}
\maketitle
%
\begin{abstract}
\noindent
In the framework of the ESPRIT project 28620 ``TIRAN'' (tailorable fault tolerance 
frameworks for embedded applications), a toolset of error detection, isolation,
and recovery components is being designed to serve as a basic means for orchestrating
application-level 
fault tolerance.  These tools will be used either as stand-alone components 
or as the peripheral components of a distributed application, that we
call ``the backbone''. The backbone is to run in the background of the 
user application. Its objectives include (1) gathering and maintaining 
error detection information produced by TIRAN components like
watchdog timers, trap handlers, or by external detection services
working at kernel or driver level, and (2) using this information
at error recovery time. In particular, those TIRAN tools related to error
detection and fault masking will forward their deductions to the backbone
that, in turn, will make use of this information to orchestrate error recovery,
requesting recovery and reconfiguration actions to those tools
related to error isolation and recovery.
Clearly a key point in this approach is guaranteeing that the backbone itself
tolerates internal and external faults. In this article
we describe one of the means that are used within the TIRAN backbone
to fulfill this goal: a distributed algorithm for tolerating crash failures
triggered by faults affecting at most all but one of the components of
the backbone or at most all but one of the nodes of the system.
We call this the algorithm of mutual suspicion.
%
\end{abstract}
%
\Section{Introduction}
In the framework of the ESPRIT project 28620 ``TIRAN''~\cite{BDDC99+},
a toolset of error detection, isolation, and recovery components is being
developed
to serve as a basic means for orchestrating application-level software 
fault tolerance. The basic components of this toolset can be considered
as ready-made software tools that the developer has to embed into 
his/her application so to enhance its dependability. These tools 
include, e.g., watchdog timers, trap handlers, local and distributed
voting tools.

The main difference between TIRAN and other
libraries with similar purposes, e.g., ISIS~\cite{Bir85} or HATS~\cite{HuKi95},
is the adoption of a special component, located between the basic
toolset and the user application. This entity is
transparently replicated on each node
of the system, to keep track of events originating in the basic layer
(e.g., a missing heartbeat from a task guarded by a watchdog)
or in the user application (e.g., the spawning of a new task),
and to allow the orchestration of system-wide error recovery
and reconfiguration. We call this component ``the backbone''.

In order to perform these tasks, the backbone hooks to each 
active instance of the basic tools and is transparently informed of 
any error detection or fault masking event taking place in the
system. Similarly, it also hooks to a library of basic services.
This library includes, among others, functions for remote
communication, for task creation and management, and to access
the local hardware clock. These functions are instrumented so
to transparently forward to the backbone notifications
of events like the creation or the termination of a thread.
Special low-level services at kernel or at driver-level are also
hooked to the backbone---for example, on a custom board
based on Analog Devices ADSP-21020 DSP and on IEEE 1355-compliant
communication chips~\cite{ieee1355}, communication
faults are transparently notified to the backbone by driver-level tools.
Doing this, an information stream flows on each node
from different abstract layers to the local component of
the backbone. This component maintains and updates
this information in the form of a system database, also 
replicating it on different nodes. 

\begin{figure}[t]
\centerline{\psfig{figure=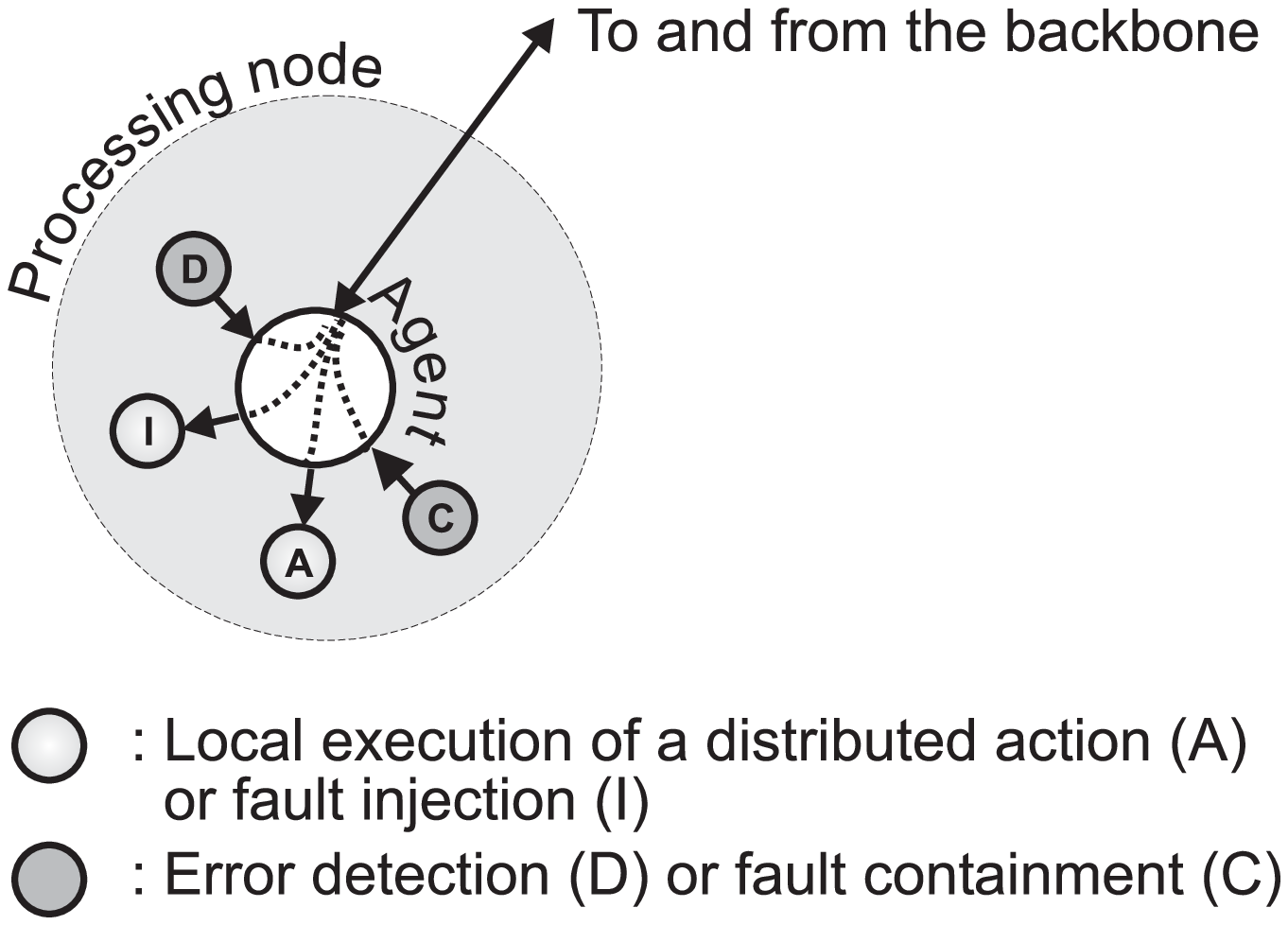,height=6.2cm}}
\caption{Each processing node of the system hosts one agent of the
  backbone. This component is the intermediary of the backbone on that
  node. In particular, it gathers information from the TIRAN tools
  for error detection and fault containment (grey circles) and forwards
  requests to those tools managing error containment and recovery (light
  grey circles). These latter execute recovery actions and possibly, at
  test time, fault injection requests.}
\label{Fig:Interactions}
\end{figure}

Whenever an error is detected or a fault is masked,
those TIRAN tools related to error detection and fault masking
forward their deductions to the backbone that, in turn,
makes use of this information to manage error recovery,
requesting recovery and reconfiguration actions to those tools
related to error isolation and recovery (see Fig.~\ref{Fig:Interactions}).

The specification of which actions to take is to be supplied 
by the user in the form of a ``recovery script'', 
a sort of ancillary application context devoted to error recovery concerns,
consisting of a number of guarded commands: the execution of blocks of
basic recovery actions, e.g., restarting a group of tasks, rebooting a node
and so forth, is then subject to the evaluation of boolean clauses based on
the current contents of the system database---for further
information on this subject see~\cite{DeDL99a,GTDR99}.

Clearly a key point in this scheme is guaranteeing that the backbone
itself tolerates internal as well external faults. In this article
we describe one of the means that have been designed within the TIRAN backbone
to fulfill this goal: a distributed algorithm for tolerating crash failures
triggered by faults affecting at most all but one of the components of
the backbone. We call this the algorithm of mutual suspicion.
As in~\cite{Cri95b}, we assume a timed asynchronous distributed
system model~\cite{CrFe99}.
Furthermore, we assume the availability of asynchronous communication
means. No atomic broadcast primitive is required.

\Section{Basic assumptions}
The target system is a distributed system consisting of $n$ nodes ($n\ge1$).
Nodes are assumed to be labeled
with a unique number in  $\{0,\dots,n-1\}$. The backbone is a  distributed
application consisting of $n$ triplets of tasks:
	  \[ (\hbox{\taskd}, \hbox{\taski}, \hbox{\taskr}). \]
Basically \taskd{} bears its name from the fact that it deals with the system
\emph{d\/}atabase of the backbone, while \taski{} manages ``I'm alive'' signals,
and \taskr{} deals with error \emph{r\/}ecovery (see further on).
We call ``agent''  any such triplet. At initialisation time, on each node
there is exactly one agent, identified
through the label of the node where it runs on.
For any $0\le k<n$, we  call ``$t[k]$''  task  $t$ of
agent $k$. Agents  play two  roles: coordinator
(also called manager)  and assistant (also called backup agent or
simply ``backup'').  In the initial, correct  state there is
just one coordinator. The  choice of which node should host the coordinator,
as well as system configuration and node labeling is  done at compile time
through a configuration script.

\begin{figure}[t]
\hskip-0.7cm\hbox{\psfig{figure=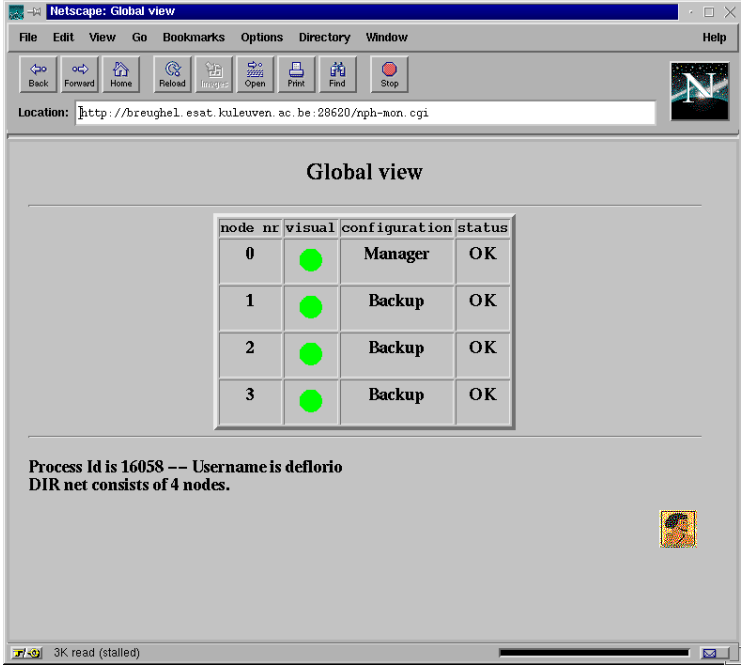,width=9.8cm}}
\caption{A Netscape browser offers a global view of the TIRAN backbone:
number, role, and state of each component is displayed.
``DIR net'' is a nickname for the backbone.
``OK'' means that no faults have been detected.
When the user selects a component, further information related to that
component is displayed. The small icon on bottom links to a page with
information related to error recovery.}
\label{fig:5}
\end{figure}

Figure~\ref{fig:5} displays a backbone running on a four node system
(a Parsytec Xplorer MIMD engine based on PowerPC microprocessors).
In this case, node zero hosts the coordinator while nodes 1--3
execute as assistants. In the portrayed situation no errors
have been detected, and this is rendered with green circles and
a status label equal to``OK''.

Each  agent, be it a coordinator or an assistant, executes a number of
common tasks.  Among  these  tasks  we  have:

\begin{enumerate}
  \item Interfacing the instances of the basic tools of the framework---this is to be
     carried out by \taskd.
  \item Organizing/maintaining data gathered from the instances---also
     specific of \taskd.
  \item Error recovery and reconfiguration management (\taskr).
  \item Self-check. This takes place through a distributed algorithm that is
     executed by \taskd{} and \taski{} in all agents in the case that $n>1$.
\end{enumerate}

This article does not cover points 1--3; in particular points 1 and 2
are dealt with as reported in~\cite{DDLB99a}, while the issue of
error recovery and reconfiguration management is described in~\cite{DeDL99a}.
In the following, we describe the algorithm mentioned at point 4. As the key
point of this algorithm is the fact that the coordinator and 
assistants mutually question their state, we call this
the algorithm of mutual suspicion (AMS).

\Section{The algorithm of mutual suspicion}
This Section describes AMS. Simulation and testing show that
AMS is capable of tolerating crash failures
of up to $n-1$ agents (some or all of which may by caused by a node crash).
This implies fail-stop behaviour, that can be reached in hardware, e.g.,
by architectures based on duplication and comparison~\cite{Pra96},
and coupled with other
techniques like, e.g., control flow monitoring~\cite{Sc87}
or diversification and comparison~\cite{Pow97a}.
If a coordinator or its node crash, a non-crashed assistant becomes
coordinator. Whenever a crashed agent is restarted, possibly after
a node reboot, it is inserted again in the backbone as an assistant.
Let us first sketch the structure of AMS. Let $m$ be the node where
the coordinator first runs on. In short:

\begin{figure}[t]
\centerline{\psfig{figure=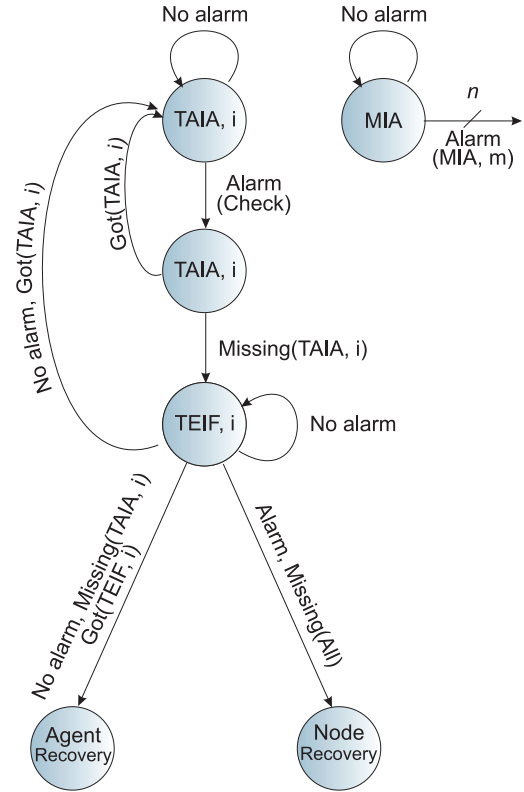,width=7.0cm}}
\caption{A representation of the algorithm of the manager.}
\label{fig:1}
\end{figure}

\begin{figure}[t]
\centerline{\psfig{figure=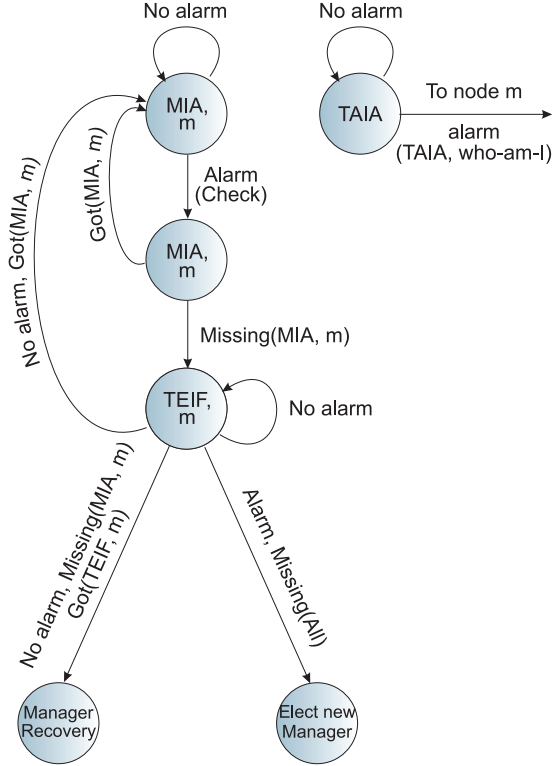,width=7.5cm}}
\caption{A representation of the algorithm of the assistant.}
\label{fig:2}
\end{figure}

\begin{figure}[t]
\centerline{\psfig{figure=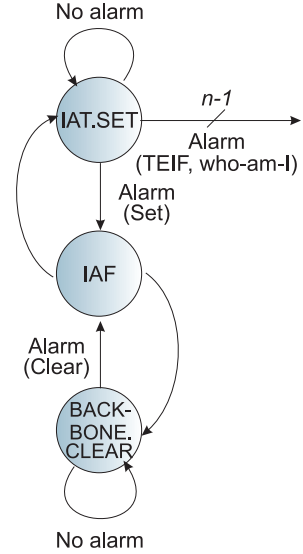,width=4.0cm}}
\caption{A representation of the conjoint action of \taski{} and \taskd{} on
the ``I'm alive'' flag.}
\label{fig:3}
\end{figure}

\begin{itemize}
   \item The coordinator periodically broadcasts a MIA (manager is alive)
     message as part of its \taskd, the assistants periodically send the
     coordinator a TAIA (this assistant is alive) message as part of their
     \taskd.

   \item For each $0\le k<n$, \taskd$[k]$ periodically sets a flag. This
     flag is periodically cleared by \taski[$k]$.

   \item If, for any valid $j$, \taski$[j]$ finds that the flag has not been set during
     the period just elapsed, \taski[$j]$ broadcasts a TEIF (this entity is
     faulty) message.

   \item When any agent, say the agent on node $g$, does not receive any timely message from
     an other agent, say the agent on node $f$, be that a MIA or a TAIA message, then agent $g$
     enters a so-called suspicion-period by setting flag sus$[f]$. This state
     leads to three possible next states, corresponding to these events:

\begin{enumerate}
  \item Agent $g$ receives a TEIF from \taski$[f]$ within a specific time period,
     say $t$ clock ticks.
  \item Agent $g$ receives a (late) TAIA from \taskd$[f]$ within $t$ clock ticks.
  \item No message is received by agent $g$ from neither \taski$[f]$ nor \taskd$[f]$
     within $t$ clock ticks.
\end{enumerate}

     State 1  corresponds to deduction ``agent on node $f$  has crashed, 
     though node $f$ is still operational''. State 2 translates into deduction 
     ``both agent on node $f$ and node  $f$ are  operational, though for  
     some reason agent  $f$ or its communication means have been
     slowed down''. State 3 is the detection of a crash failure for node $f$.
     These deductions  lead to actions  aiming at recovering agent  $f$ or (if
     possible)  the  whole  node $f$,  possibly  electing  a new  coordinator.
     In the present version of AMS, the election algorithm is simply carried out
     assuming the next coordinator to be the assistant on node $m+1 \,\mod\, n$.
\end{itemize}

\begin{figure}
\psfig{figure=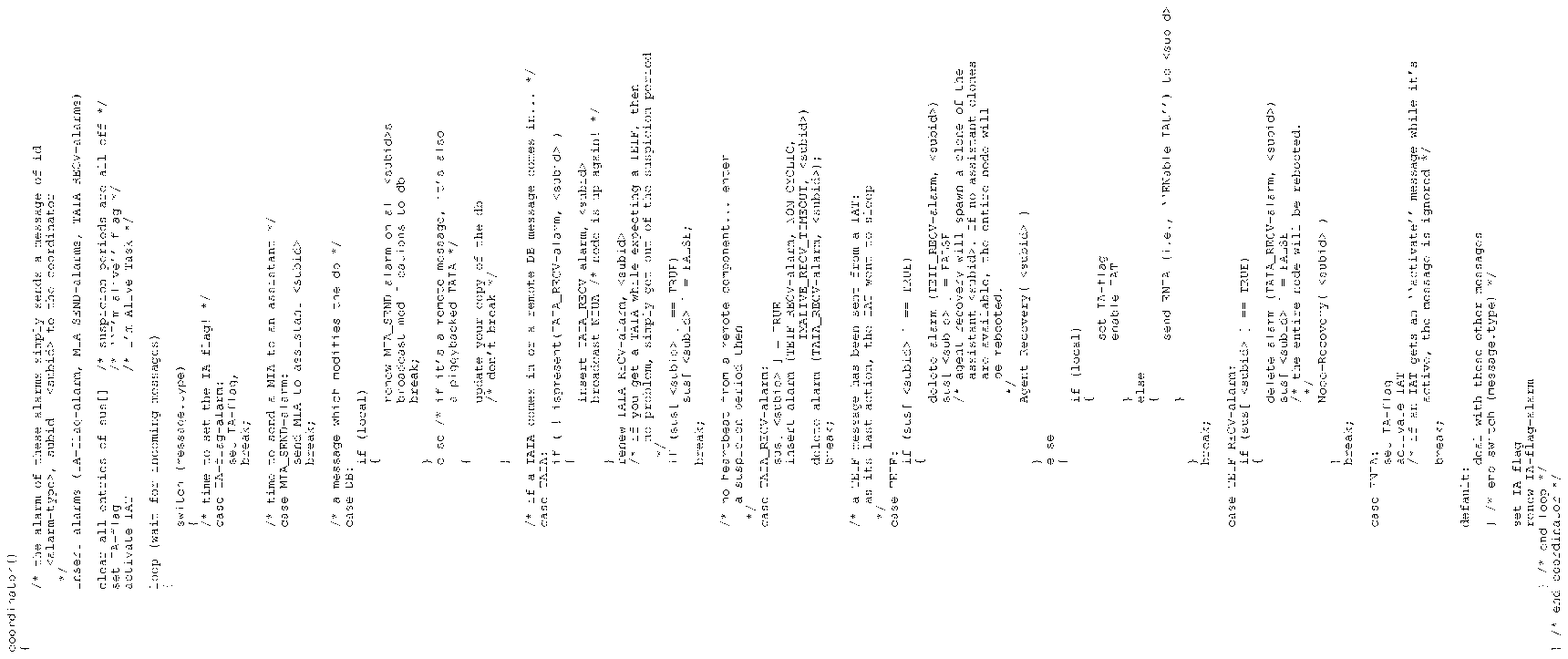,width=20.4cm,angle=-90}
\caption{Pseudo-code of the coordinator.}
\label{pseudoc}
\end{figure}

As compliant to the timed asynchronous distributed system model,
we assume the presence of an alarm manager task (\taska) on each node of the
system,  spawned at  initialization time  by the  agent. \Taska{} is  used to
translate  time-related clauses, e.g.,  ``$t$  clock ticks have  elapsed'', into
message arrivals. Let  us call \taska$[j]$ the \taska{} running on node $j$,  $0\le j<n$. \Taska{}
may be represented as a function

                                  \[ a: C \to M \]
\noindent
such that, for any  time-related clause $c\in C$,
\[ a(c) = \hbox{message ``clause $c$ has elapsed'' } \in M. \]

\Taska$[j]$ monitors a set of time-related
clauses and, each time  one of them occurs, say clause $c$, it sends \taskd$[j]$
message $a(c)$. Messages are sent via asynchronous primitives based on mailboxes.

In  particular, the  coordinator, which  we assume  to run
initially on node  $m$, instructs its \taska{} so  that the following clauses be
managed:

   \begin{itemize}
   \item (MIA\_SEND, $j$, MIA\_SEND\UNDSCR{}TIMEOUT), $j$ different from $m$: every
     MIA\_SEND\UNDSCR{}TIMEOUT clock ticks, a message of type (MIA\_SEND, $j$) should be
     sent to \taskd$[m]$, i.e., \taskd{} on the current node. This latter will
     respond to such event by sending each \taskd$[j]$ a MIA (manager is
     alive) message.

   \item (TAIA\_RECV, $j$, TAIA\_RECV\UNDSCR{}TIMEOUT) for each $j$ different from $m$: every
     TAIA\_RECV\UNDSCR{}TIMEOUT clock ticks at most, a message of type TAIA is to be
     received from \taskd$[j]$. The arrival of such a message or of any other
     ``sign of life'' from \taskd$[j]$ translates also in renewing the
     corresponding alarm. On the other hand, the arrival of a message of
     type (TAIA\_RECV, $k$), $k$ different
     from $m$, sent by \taska$[m]$, warns \taskd$[m]$
     that assistant on node $k$ sent no sign of life throughout
     the TAIA\_RECV\UNDSCR{}TIMEOUT-clock-tick period just elapsed. This makes 
     \taskd$[m]$ set its flag sus$[k]$.

   \item (I'M\_ALIVE\_SET, $m$, I'M\_ALIVE\_SET\UNDSCR{}TIMEOUT): every I'M\_ALIVE\_SET\UNDSCR{}TIMEOUT
     clock ticks, \taska$[m]$ sends \taskd$[m]$ an I'M\_ALIVE\_SET message. As a
     response to this, \taskd$[m]$ sets the ``I'm alive'' flag, a memory variable
     that is shared between tasks of type $\mathcal D$ and $\mathcal I$.
   \end{itemize}

Furthermore, whenever  flag sus$[k]$ is set,  for any $k$ different  from $m$, the
following clause is sent to \taska{} for being managed:

\begin{itemize}
   \item (TEIF\UNDSCR{}RECV, $k$, TEIF\UNDSCR{}RECV\UNDSCR{}TIMEOUT): this clause simply 
     asks \taska$[m]$ to
     schedule the sending of message (TEIF\UNDSCR{}RECV, $k$) to \taskd$[m]$ after
     TEIF\UNDSCR{}RECV\UNDSCR{}TIMEOUT clock ticks. This action is canceled should a
     late TAIA message arrive to \taskd$[m]$ from task $k$, or should a TEIF
     message from \taski$[k]$ arrive instead. In the first case, sus$[k]$ is
     cleared and (possibly empty) actions corresponding to a slowed down
     \taskd$[k]$ are taken. In the latter case, \taskd$[k]$ is assumed to have
     crashed, its clauses are removed from the list of those managed by 
     \taska$[m]$, and flag sus$[k]$ is cleared. It is assumed that \taski$[k]$ will
     take care in this case of reviving \taskd$[k]$. Any future sign of life
     from \taskd$[k]$ is assumed to mean that \taskd$[k]$ is back in operation.
     In such a case \taskd$[k]$ would then be re-entered in the list of
     operational assistants, and \taskd$[m]$ would then request \taska$[m]$ to include again an
     alarm of type (MIA\_SEND, $k$, MIA\_SEND\UNDSCR{}TIMEOUT) and an alarm of type
     (TAIA\_RECV, $k$, TAIA\_RECV\UNDSCR{}TIMEOUT) in its list. If a (TEIF\UNDSCR{}RECV, $k$)
     message reaches \taskd$[m]$, the entire node $k$ is assumed to have
     crashed. Node recovery may start at this point, if available, or
     a warning message should be sent to an external operator so that, e.g.,
     node $k$ be rebooted.
\end{itemize}


Similarly any  assistant, say the one on node $k$,  instructs its \taska{} so
that the following clauses be managed:

\begin{itemize}
\item (TAIA\_SEND, $m$, TAIA\_SEND\UNDSCR{}TIMEOUT): every TAIA\_SEND\UNDSCR{}TIMEOUT clock ticks,
     a message of type (TAIA\_SEND, $m$) is to be sent to \taskd$[k]$, i.e.,
     \taskd{} on the current node. This latter will respond to such event by
     sending \taskd$[m]$ (i.e., the manager) a TAIA (this agent is alive)
     message. Should a data message be sent to the manager in the middle of
     TAIA\_SEND\UNDSCR{}TIMEOUT-clock-tick period, alarm (TAIA\_SEND, $m$,
     TAIA\_SEND\UNDSCR{}TIMEOUT) is renewed. This may happen for instance because one
     of the basic TIRAN tools for error detection
     reports an event to \taskd$[k]$. Such event must be sent
     to the manager for it to update its database. In this case we say that
     the TAIA message is sent in piggybacking with the event notification
     message.

\begin{figure}
\psfig{figure=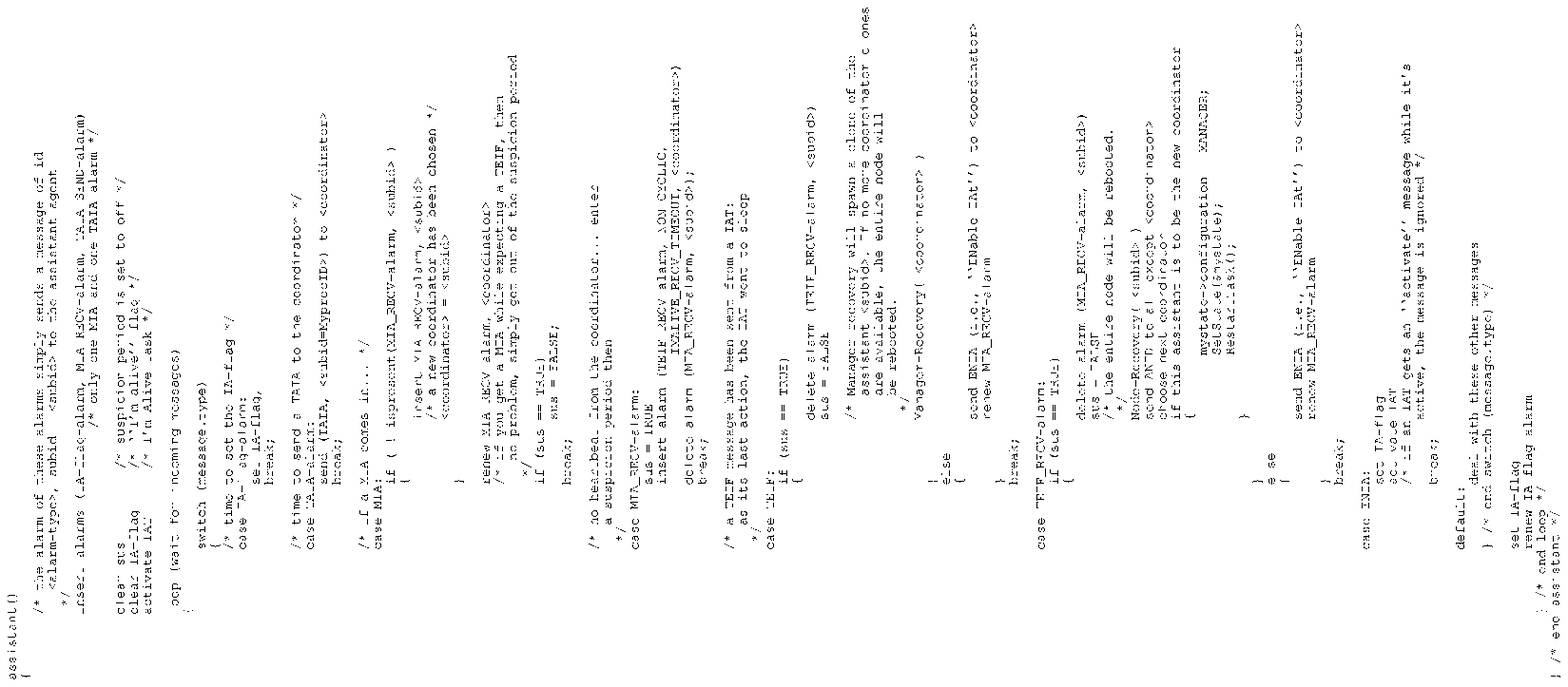,width=20.4cm,angle=-90}
\caption{Pseudo-code of the assistant.}
\label{pseudoa}
\end{figure}

\item (\hbox{MIA\_\hskip1pt RECV}, $m$, MIA\_RECV\UNDSCR{}TIMEOUT): every MIA\_RECV\UNDSCR{}TIMEOUT clock ticks at
     most, a message of type MIA is to be received from \taskd$[m]$, i.e., the
     manager. The arrival of such a message or of any other ``sign of life''
     from the manager translates also in renewing the corresponding alarm.
     If a message of type (MIA\_RECV, $m$), is received from \taskd$[k]$ and sent
     by \taska$[k]$, this means that no sign of life has been received from
     the manager throughout the MIA\_RECV\UNDSCR{}TIMEOUT-clock-tick period just
     elapsed. This makes \taskd$[k]$ set flag sus$[m]$.

\item (I'M\_ALIVE\_SET, $k$, I'M\_ALIVE\_SET\UNDSCR{}TIMEOUT): every I'M\_ALIVE\_SET\UNDSCR{}TIMEOUT
     clock ticks, \taska$[k]$ sends \taskd$[k]$ an I'M\_ALIVE\_SET message. As a
     response to this, \taskd$[k]$ sets the ``I'm alive'' flag.
   \end{itemize}

Furthermore, whenever  flag sus$[m]$ is  set, the following clause  is sent to
\taska$[k]$ for being managed:

\begin{itemize}
\item (TEIF\UNDSCR{}RECV, $m$, TEIF\UNDSCR{}RECV\UNDSCR{}TIMEOUT): this clause simply asks \taska$[k]$ to
     postpone sending message (TEIF\UNDSCR{}RECV, $k$) to \taskd$[k]$ of
     TEIF\UNDSCR{}RECV\UNDSCR{}TIMEOUT clock ticks. This action is canceled should a
     late MIA message arrive to \taskd$[k]$ from the manager, or should a TEIF
     message from \taski$[m]$ arrive instead. In the first case, sus$[m]$ is
     cleared and possibly empty actions corresponding to a slowed down
     manager are taken. In the latter case, \taskd$[m]$ is assumed to have
     crashed, its clause is removed from the list of those managed by 
     \taska$[k]$, and flag sus$[m]$ is cleared. It is assumed that \taski$[m]$ will
     take care in this case of reviving \taskd$[m]$. Any future sign of life
     from \taskd$[m]$ is assumed to mean that \taskd$[m]$ is back in operation.
     In such a case \taskd$[m]$ would be demoted to the role of assistant
     and entered in the list of operational assistants. The role of coordinator
     would then have been assigned, via an election, to an agent formerly running as assistant.
   \end{itemize}

     If a  (TEIF\UNDSCR{}RECV, $m$) message reaches \taskd$[k]$,  the entire node of the
     manager is  assumed to  have crashed. Node  recovery may start  at this
     point.   An election takes place---the next assistant (modulo $n$) is
     elected   as  new   manager.

Also  \taski{}  on each node,  say node $k$,  instructs its  \taska{} so  that the
following clause be managed:

\begin{itemize}
\item (I'M\_ALIVE\UNDSCR{}CLEAR, $k$, I'M\_ALIVE\UNDSCR{}CLEAR\UNDSCR{}TIMEOUT): every
     I'M\_ALIVE\UNDSCR{}CLEAR\UNDSCR{}TIMEOUT clock ticks \taska$[k]$ sends \taski$[k]$ an
     I'M\_ALIVE\UNDSCR{}CLEAR message. As a response to this, \taski$[k]$ clears the
     ``I'm alive'' flag.
\end{itemize}

Figures~\ref{fig:1}, \ref{fig:2}, and~\ref{fig:3} supply a
pictorial representation of this algorithm. Figure~\ref{pseudoc}
and Fig.~\ref{pseudoa} respectively
show a pseudo-code of the coordinator and of the assistant.

\SubSection{The alarm manager class}
This section briefly describes  \taska. This task makes use of a special
class  to  manage  lists of alarms~\cite{DeDL98h}.  The  class  allows the client  to
``register'' alarms, specifying alarm-ids and deadlines.

Once the  first alarm is entered,  the task managing alarms  creates a
linked-list of alarms and polls the top of the list. For each new alarm
to be  inserted, an entry in the list is found  and the list is modified
accordingly.  If the  top entry  expires, a user-defined alarm function is invoked.
This is a general mechanism that allows to associate any event with the
expiring of an alarm. In the case of the backbone, \taska{} on node $k$
sends a message to \taskd$[k]$---the same result may also be achieved by sending
an UNIX signal to \taskd$[k]$.
Special alarms are defined as ``cyclic'', i.e., they are automatically
renewed at  each new  expiration, after invoking the alarm function.
A special  function restarts  an alarm,
i.e.,  it deletes  and re-enters an  entry. It  is also  possible to
temporarily  suspend  an  alarm  and  re-enable it  afterwards.

\begin{figure*}[t]
\hskip-1.0cm\vbox{\hbox{\psfig{figure=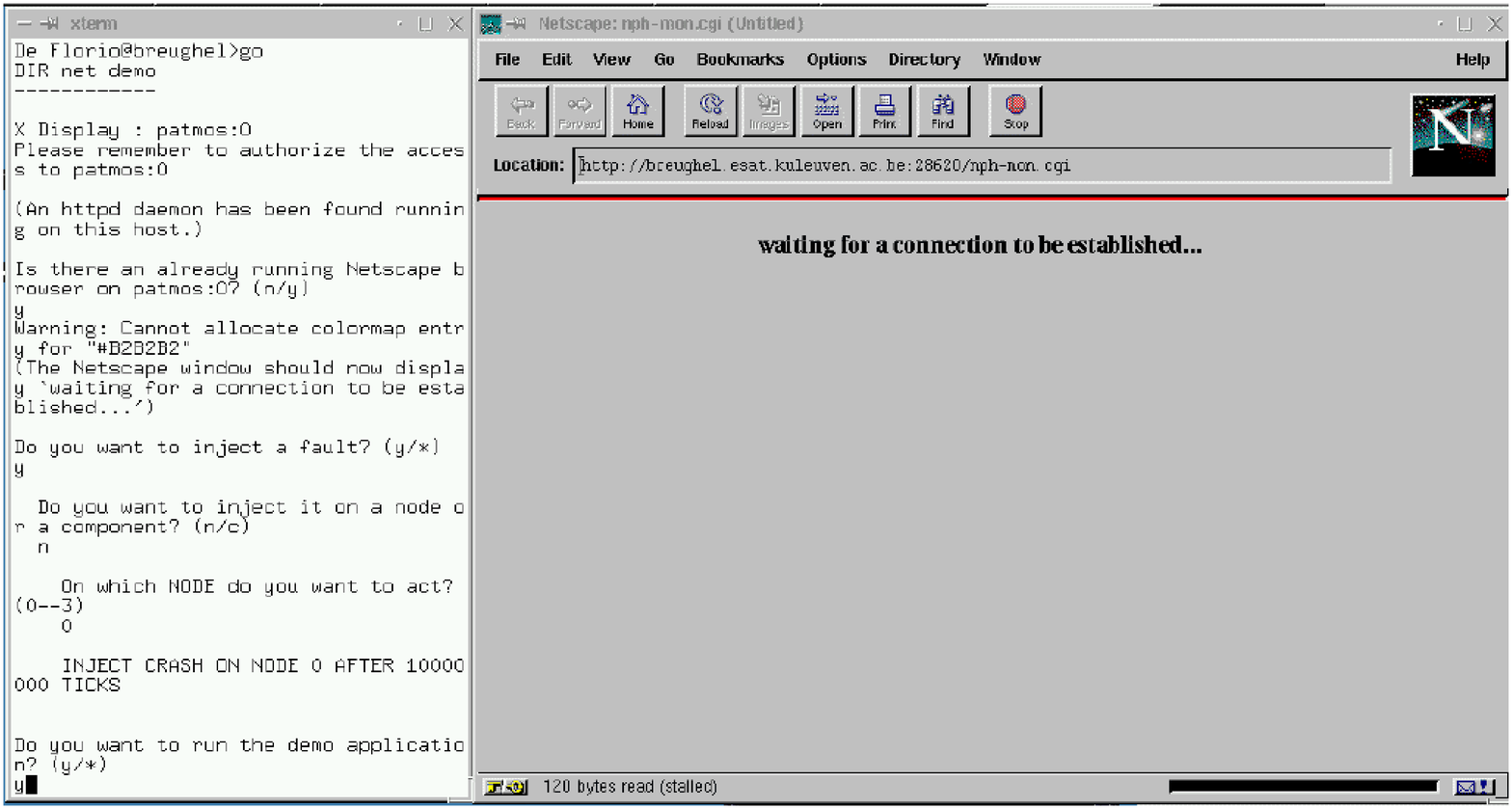,width=9.7cm}
\psfig{figure=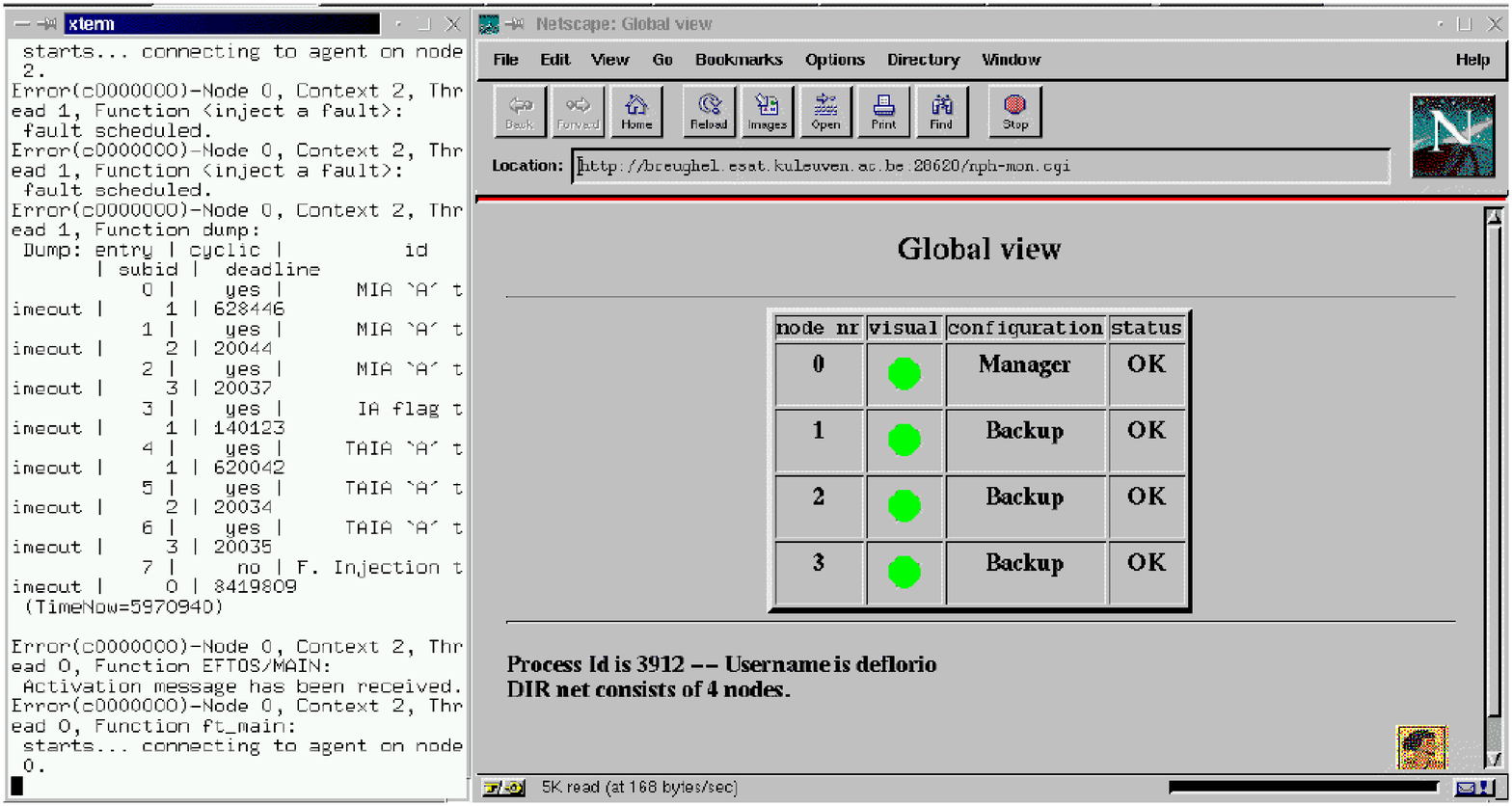,width=9.7cm}}
\hbox{\psfig{figure=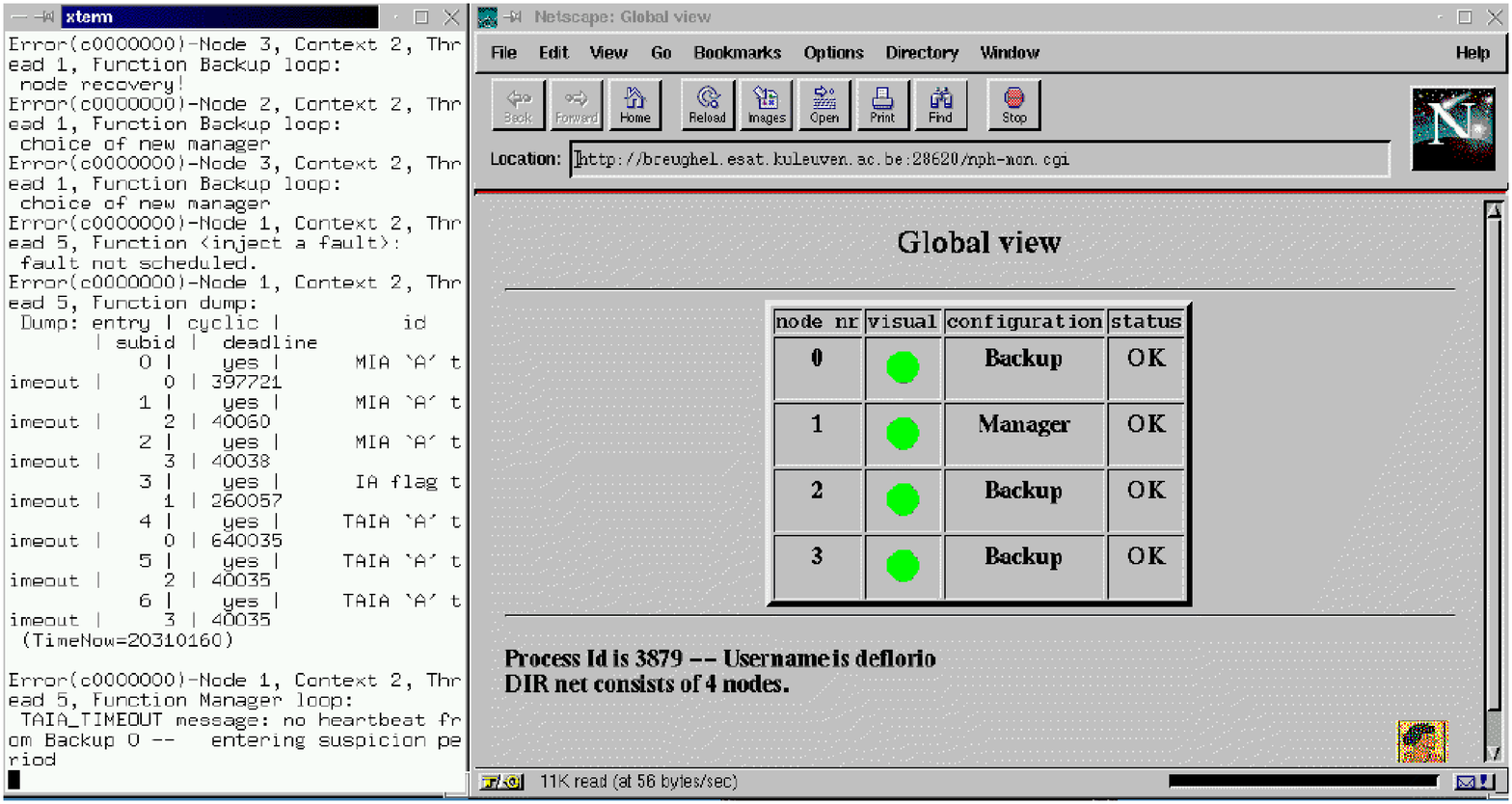,width=9.7cm}
\psfig{figure=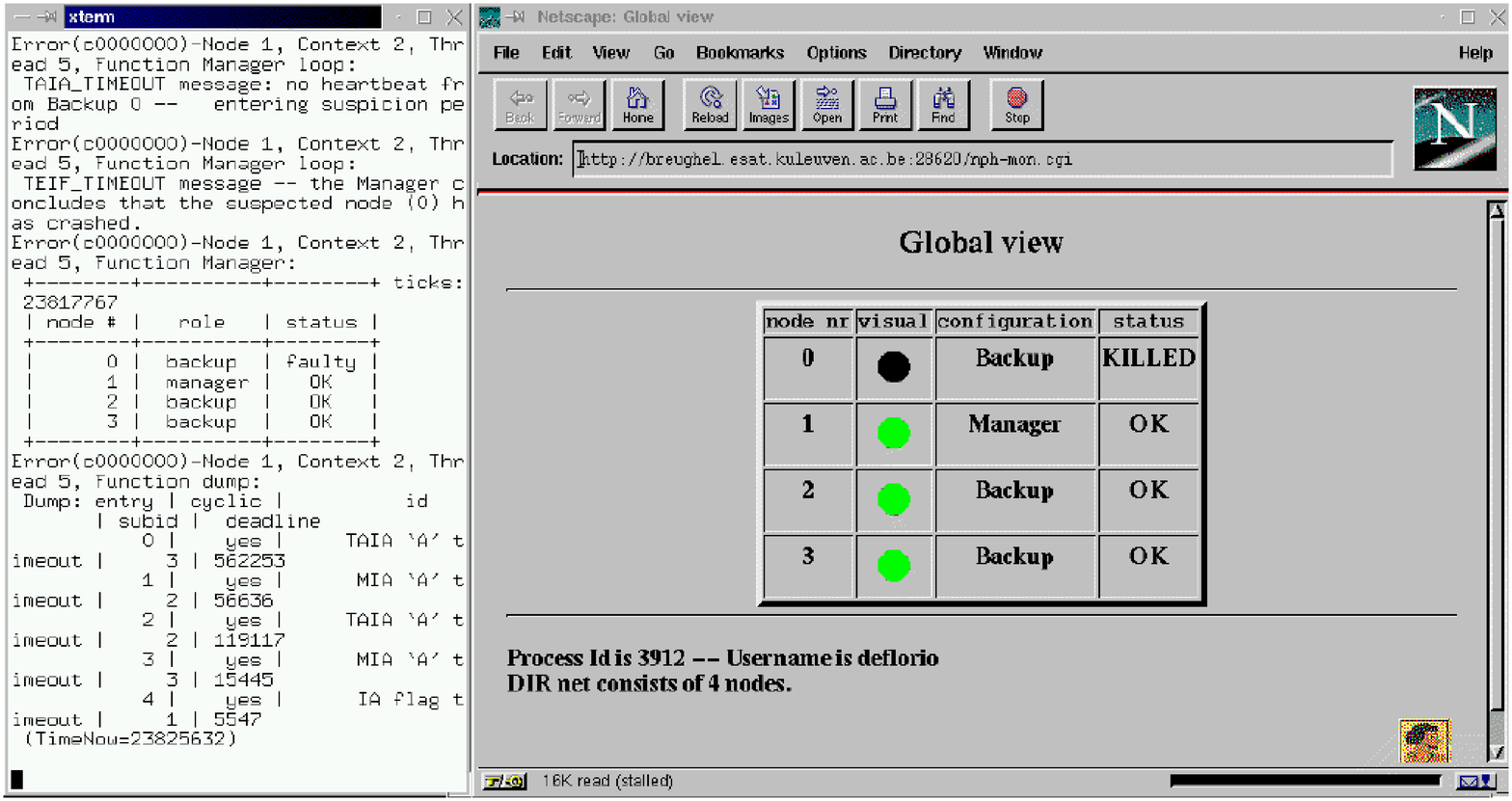,width=9.7cm}}}
\caption{A fault is injected on node 0 of a system of four nodes.
Node 0 hosts the manager of the backbone. In the top left picture the user
selects the fault to be injected and connects to a remotely controllable
Netscape browser. The top right picture shows this latter as it renders
the shape and state of the system. The textual window reports the
current contents of the list of alarms used by \taska[0].
In the bottom left picture the crash of node 0 has been
detected and a new manager has been elected. On election, the manager
expects node 0 to be back in operation after a recovery step. This 
recovery step is not performed in this case. As a consequence,
node 0 is detected as inactive and labeled as ``KILLED''.}
\label{Fig:AMS}
\end{figure*}

\begin{figure*}[t]
\centerline{\psfig{figure=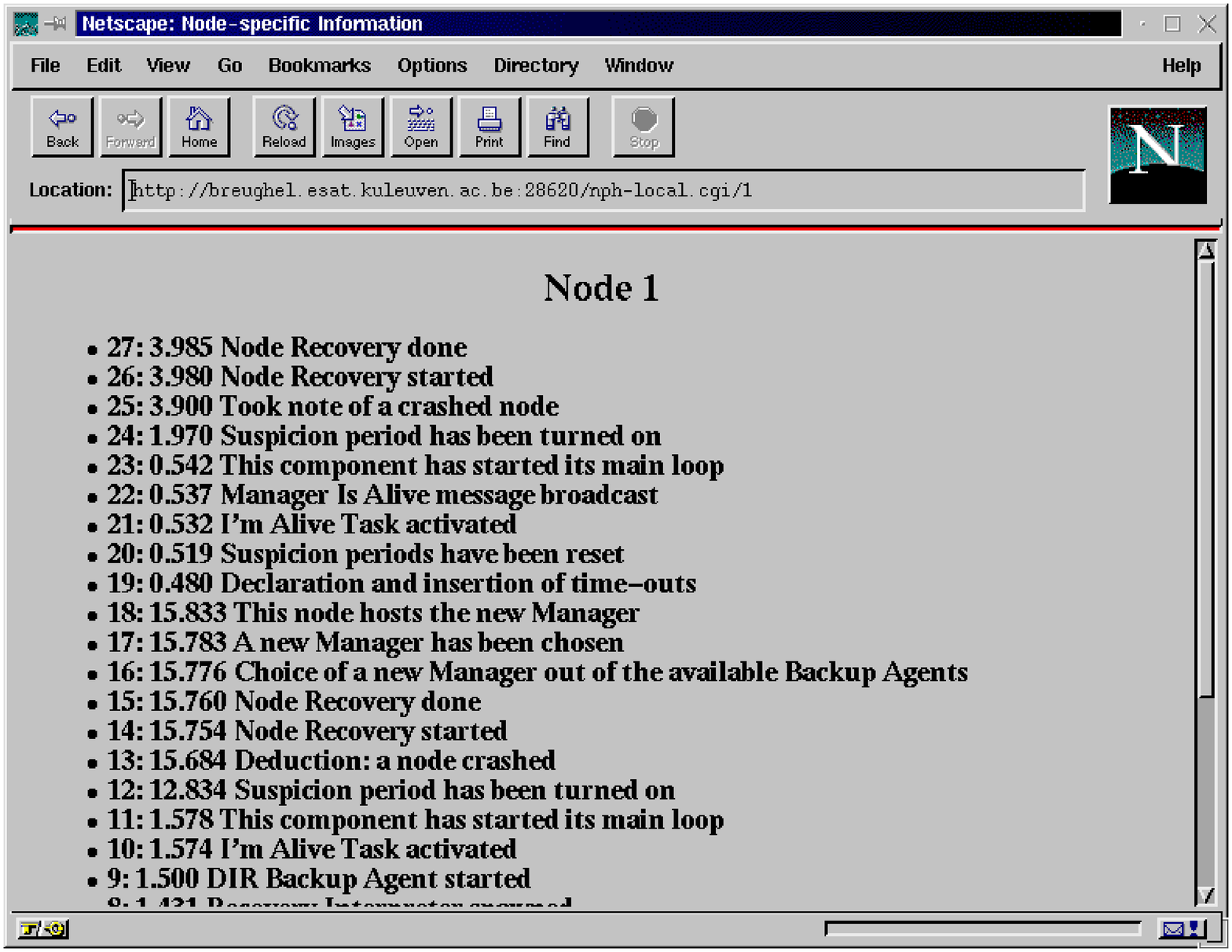,width=12.135cm}}
\caption{When the user selects a circular icon in the 
Web page of Fig.~\ref{fig:5}, the browser
replies with a listing of all the events that took place on
the corresponding node.
Here, a list of events occurred on \taskd[1] during the
experiment shown in Fig.~\ref{Fig:AMS} is displayed. Events are
labeled with an event-id and with the time of occurrence
(in seconds). Note in particular event 15, corresponding
to deduction ``a node has crashed'', and
events 16--18, in which the election of the manager takes place
and \taskd[1] takes over the role of the former manager, \taskd[0].
Restarting as manager, \taskd[1] resets the local clock.}
\label{fig:rec}
\end{figure*}

\Section{Current status and future directions}
A prototypal implementation of the TIRAN backbone is running on a Parsytec
Xplorer, a MIMD engine, using 4 PowerPC nodes. The system has been
tested and proved to be able to tolerate a number of software-injected faults,
e.g., component and node crashes (see Fig.~\ref{Fig:AMS} and
Fig.~\ref{fig:rec}).
Faults are scheduled as another class of alarms that, when triggered,
send a fault injection message to the local \taskd{} or \taski{}.
The specification of which fault to inject is read by the
backbone at initialisation time from a file called ``.faultrc''.
The user can specify fault injections by editing this file, e.g.,
as follows:
\begin{quote}
INJECT CRASH ON COMPONENT 1\\
\hspace*{33pt}      AFTER 5000000 TICKS\\
INJECT CRASH ON NODE 0\\
\hspace*{33pt}      AFTER 10000000 TICKS.
\end{quote}
The first two lines inject a crash on \taskd{}[1] after 5 seconds
from the initialisation of the backbone, the second ones
inject a system reboot of node 0 after 10 seconds.
Via fault injection it is
also possible to slow down artificially a component for a given
period. Slowed down components are temporarily and automatically
disconnected and then accepted again in the application when
their performance goes back to normal values.
Scenarios are represented
into a Netscape window where a monitoring application displays
the structure of the user application, maps the backbone roles onto the
processing nodes of the system, and constantly reports about the events
taking place in the system~\cite{DeDe98}.

This system has been recently redeveloped so to enhance its portability
and performance and to improve its resilience.
The backbone is currently being implemented
for target platforms based on Windows CE, VxWorks, and TEX~\cite{Anon97b}.
A GSPN model of the algorithm of mutual suspicion has been developed
by the University of Turin, Italy, and has been used to validate
and evaluate the system. Simulations of this model proved
the absence of deadlocks and livelocks.
Measurements of the overheads in fault free scenarios and when faults occur
will also be collected and analysed. 

\paragraph{Acknowledgments.}
This project is partly supported by an 
FWO Krediet aan Navorsers and by the ESPRIT-IV
project 28620 ``TIRAN''.

\bibliographystyle{latex8}

\end{document}